\documentclass[12pt,a4paper,aps]{article}
\usepackage{a4}
\usepackage{epsfig}
\usepackage[hmargin=2cm,vmargin=2cm,nohead]{geometry}
\begin{document}
\par
\title{The European Electricity Grid System \\ and Winter Peak Load Stress: \\
For how long can the european grid system survive the ever increasing demand 
during cold winter days?}

\author{
Michael Dittmar\thanks{e-mail:Michael.Dittmar@cern.ch},\\
Institute of Particle Physics,\\ 
ETH, 8093 Zurich, Switzerland}
\maketitle

\begin{abstract}
The rich countries of Western Europe and its citizens 
benefited during at least the last 30 years from an extraordinary stable electricity grid. 
This stability was achieved by the european grid system and a large flexible and reliable spare power 
plant capacity.  This system allowed a continuous demand growth during the past 10-20 years of up to a few \% per year. 
However, partially due to this overcapacity, no new large power plants have been completed during the past 10-15 years.
The obvious consequence is that the reliable spare capacity has been reduced and that    
a further yearly demand growth of 1-2\% for electric energy can only be achieved if new power plants will be constructed soon. 
The soon existing gap between the supply and the demand wishes of electric energy is widely acknowledged, 
but most policy and decision makers behave as if electricity shortages are always a few years away.

In reality however, the data from various European countries, provided by the UCTE, 
indicate that the system stress during peak load times and especially during particular cold winter days 
is much larger than generally assumed.
In fact, the latest UCTE data on reliable power capacity indicate that already during the 
Winter 2007/8 only a few very cold winter days in Western Europe 
could have resulted in very unpleasant supply problems during the early evening 
peak load times in France and the neighboring countries. 
These estimates of a stressed grid system 
were found to be confirmed during the evening of the 17th of December 2007, 
when new absolute electricity consumption records where observed in France, Spain and Italy. 
As the rest of the 2007/8 winter turned out to be abnormal mild, a serious test of the 
grid stability is now postponed to the next winter. 

An analysis of the potential to build and operate new power plants demonstrates that no 
short term solution, which allows a further continuous 1-2\% consumption growth seems to exist.  
Thus, without any action from the various governments in Europe to reduce the electricity demand 
in a serious way, the risk of blackouts will increase dramatically in the near future.
Instead of simply waiting for the blackout disaster to happen, a     
few simple short term political actions are suggested. If such actions are taken well in advance, 
the risks for peak load related blackouts might be reduced sufficiently.

For those who assume that no coordinated action will be taken and that 
``normal" cold severe winter days will happen eventually, some simple suggestions 
might reduce the most nasty personal consequences of blackouts. 
\end{abstract}


\newpage
\section{Introduction}

Electric energy is known to be the most flexible and most useful form of energy 
for stationary needs of a highly industrialized society. Today, almost nothing  
in our houses, in the manufactureing and servicing industry
functions without electric energy.  
The availability of abundant and relatively cheap electric energy during at least the last 30 years 
in Western Europe and other highly industrialized countries has lead a large number of people
almost to the believe and behavior that this wonderful energy form simply comes out of the plug.
However, it is also well known that sufficient electric power generation is required to satisfy these needs
and that a continuous demand growth for electric energy can not be satisfied without 
building new power plants. 
Furthermore, it is generally assumed that local demand fluctuations 
and possible short term failures of some power plants require a spare capacity of at least 5-10\% in order to
prevent blackouts, e.g. large scale power cuts \cite{powercuts}. 
Obviously, power cuts, which can last from minutes to hours and even days 
might have very unpleasant and expensive consequences for individuals and for the society\cite{examples}.
Even though we know that power cuts are always possible, only 
for exceptional security reasons, like in hospitals, some short time and small scale backup diesel power generator capacity exists. 

Sometimes it is acknowledged that the existing power plants in Europe are  
aging and that uncurbed demand growth can only be satisfied if 
new power plants will be constructed soon \cite{Axpo2020}. 
The alternative to reduce the demand, especially during peak load times, 
in order to prevent blackouts, is so far considered by most as an unacceptable 
and impossible exotic option, which exists only for poor countries.

Unfortunately, discussions at all levels about future energy use and new power plants are fought mostly 
on an emotional basis about the pro and contra of various types of power plants.
The issues are further complicated by the lobbying of various economic interest groups
who care much more about their own short term benefits than about 
the stability of the electric grid.
Despite these philosophical and political fights, essentially all participants assume that demand growth 
and thus economic growth must continue and that 
theoretically all power plant options are only a question of the price 
and not of the resource availability. 
Furthermore, these discussions usually ignore also the importance of a flexible power production grid system, 
which can rapidly react to demand fluctuations and to provide sufficient power for the 
peak load. 

As a consequence of these mostly fruitless discussions and the ``Yes and No" statements with respect 
to the needs to reduce CO$_{2}$ production or to nuclear power plants, 
no european wide decisions have been taken so far.  Consequently 
serious shortages of electric energy become more and more likely during the coming years.
Some documents from the Union for the Co-ordination of Transmission of Electricity 
(UCTE) in Europe discuss how critical this situation might become during the coming years. 
In fact, the more detailed winter 2007/8 outlook report \cite{UCTEWinter07} shows 
that only a few real cold winter days during early 2008 would have been enough to 
create a real stress for the european electric energy supply system. 
Surprisingly, this nearby critical serious situation has so far been overlooked by 
the media and essentially by all politicians. 

To make things worse, the supply situation during winter peak load times seems to be even 
more critical than assumed in the UCTE document. A few days, a few degrees colder than the average
temperature for this period during December 2007, resulted in four days of 
unprecedented record consumption in France during the early evenings \cite{francerecord}.
Especially remarkable was the evening of the 17th of December, where in addition to France, 
new record consumptions were observed also in Italy and Spain \cite{lemonde}.
The demand in France, usually considered to have sufficient spare and export capacity, was so high that  
electric power, equivalent of 1.6 GWe, were imported from Germany and Switzerland to satisfy demand and to avoid power outages. Combining these unexpected early record consumptions with 
the ``well" known stress situation expected during a few cold winter days one should 
conclude that the situation is already now much more critical than usually assumed.

The most critical situation for the european grid system will happen 
if a cold wave will be observed in France and simultaneously in the neighboring countries.
The risk is especially high as the stability of the supply in some neighboring countries depend 
already on the potentially available spare electric power capacity from France.
For example Switzerland has some import agreement of about 2-3 GWe with France. 
France is also exporting electric energy to other countries, like Belgium and Great Britain. 
For sure, additional unplanned outages of aging power plants,
like the 2007/08 ones of two large nuclear power plants Brunsb\"uttel and Kr\"ummel in Germany,
will make the situation worse.  

In the following, we summarize the two recent 
UCTE reports, which discuss the  
system adequacy for the next 10-15 years \cite{ucte1} in section 2 
and the particular Winter 2007/8 
situation \cite{UCTEWinter07} in section 3. 

As it turns out, the winter supply situation has already reached a critical level.  
Therefore, it seems, that the time has come to ask the
various governments for some actions about reducing the risk of blackouts.
In section 4 we propose simple measures which 
could reduce the current blackout risks by asking for peak load demand reductions.
As it is unlikely that governments will take such actions during the next few years, 
some additional suggestions might help individuals, who are concerned about the perhaps 
nasty consequences of blackouts. 

\section{The European Electricity Grid and its needs} 

A highly sophisticated and efficient european wide electricity exchange system 
has been constructed during the last 50 years. It allows the ``average" European
citizen to benefit on average from about 7000 kWh per year, a factor of three more 
kWh electric energy as the current per capita world average. 
Of course, depending on the individual financial ``richness" of a person and country, 
the electric energy consumption per person within the different countries 
varies by large factors.  Furthermore, regional differences in Europe are also very large.
For example one finds that an average person in France, Germany or Switzerland 
uses about 7000 kWh per year, while the average person in Poland, Hungary, Greece
and Portugal uses about a factor of two less. Development towards ``better life 
quality" within Europe and within the different countries is usually identified 
with higher and higher consumption of many items, including electric energy.
Such a 1-2\% growth of electric energy consumption was observed during
the past 10-20 years within the  European grid system.  
Regional consumption patterns and shortages due to seasonal variations 
have been balanced succesfully during many years within this system.
This system is coordinated by the UCTE which defines itself as \cite{defucte}:

{\it The "Union for the Co-ordination of Transmission of Electricity" is the association of transmission system operators in continental Europe, providing a reliable market base by efficient and secure electric "power highways". } 

The UCTE publishes on a regular basis reports about the grid performance, covering the past  
months and years and the different european countries and regions \cite{uctereports}. In addition,
two regular yearly reports about the so called system adequacy outlook for the next 10-15 years
\cite{ucte1} and the situation during the 
high demand winter period \cite{UCTEWinter07} are of particular 
interest for the analysis described in the following.  

Especially the media release for the latest ``System Adequacy Forecast 2008-2020", 
presented on the 14th of January 2008, contains some remarkable statements:

{\it 
``New firm investment decisions seem to be necessary to maintain the 
generating capacity above the adequacy reference margin after 2015. This 
may be also later, provided that the potential for load reduction {\bf via demand 
side management is reliably available} to achieve power balance."
}

\noindent
and
 
{\it ``Investments already decided are nevertheless not sufficient to face the 
expected decrease of available generation between 2015 and 2020. 
Development of renewable sources, especially wind generation, provides 
new generation capacities but their contribution to system adequacy is only 
partial due to intermittency. The need for further investment decisions in 
new generation capacities for 2020 represents about 50 GW."}

This huge capacity number of 50 GWe (e for electric) corresponds almost to the equivalent of todays french nuclear power 
plants, or to about 30 new big nuclear ``EPR"  reactors or about 2.5 times the currently installed 
wind power system in Germany, about 20 GWe. As will be explained later, 
the ``installed" wind power number should be multiplied by another factor of 4-5 to compensate for the small capacity factor.

The above two quotes indicate that some Europe wide decisions, either 
to build new capacities or to reduce the consumption of electric energy, must be taken soon. 

The  ``Winter Outlook Report 2007", \cite{UCTEWinter07}, about the system adequacy during the 
peak load demand contains more detailed and not really reassuring statements backed up by 
data from the different countries. The following quotes from this report 
seem to describe the real situation well: 

{\it \bf{``No particular risk of power shortage is expected for the winter 2007-2008 under normal conditions."}}

\noindent
and  

{\it \bf{``Conversely, under severe conditions, due mainly to low 
temperature or unfavourable hydro-conditions, the power systems might be 
stressed, especially when the same periods are critical for neighbouring 
countries as well."}} \\

\noindent
The list of countries with ``critical" conditions in this report is already impressive! \\

{\it ``Only a few countries may depend on imports from their neighbours in some specific periods: 
Belgium, Portugal, Greece, Romania, Slovenia and the Slovak Republic. 
Finland, Former Yugoslav Republic of Macedonia, Serbia and Latvia need imports to reach 
adequate margins."}

\noindent
The report assumes further that: \\

{\it ``In these cases the transmission capacities allow for the required imports if intact."} and  \\ 
{\it \bf{``(The) Availability of the 
simultaneous export capability in the neighbouring countries or regions has not been analysed.}" 
}

\noindent \\
Under so called ``severe" conditions the list becomes even longer: \\

{\it
``In such periods, unfavourable conditions could reduce the export capabilities from exporting 
countries and could lead to tight situations at the regional level in Central Western Europe 
(Great Britain, France, Belgium and The Netherlands) and South Eastern Europe (Former Yugoslav 
Republic of Macedonia, Serbia, Greece, Romania). Among the Nordic countries, Finland, Sweden 
and Eastern Denmark will have a deficit under severe conditions, but the total Nordic generation 
capability exceeds the simultaneous peak demand. 
In addition Spain, Austria, Italy and Hungary stress the risks linked to the gas market. "}

For some unknown reasons, even though these statements sound already very alarming, 
the media and governments have evidently failed to transmit this message to a 
wider public audience. 

Before discussing the particular critical winter peak load situation, a description of how
electric energy is provided within the european grid system and what can be expected for the coming 10-15 years.

\subsection{The european electric grid system adequacy situation now and in the near future}

The total reliable available capacity for January and July 2008 in the UCTE countries 
has been estimated to be roughly 457.1 GWe and 414.3 GWe respectively \cite{ucte1}.
These numbers are up to 30\% smaller than the so called ``Net Generating Capacity", 
which is given in the same report as 643.5 GWe and 649 GWe for January and July 2008.
Out of the total net capacity, nuclear power provides 112 GWe, fossil fueled power plants 
333 GWe and hydro power 136 GWe. Other renewable energy sources, mainly windpower, 
contribute now a potential of about 62 GWe.

The overall production of electric energy in the UCTE countries 
in 2006 was  2584.6 TWh, out of which 802 TWh came from nuclear, 306 TWh 
from hydro and 1360 TWh from fossil fueled power plants\cite{ucte3}. 
Wind power and other renewable power plants contributed about 117 TWh. 
Comparing the capacity of the different types of 
power plants with the annual produced electric energy one finds that the so called capacity 
factor, the real produced number of kWh per year divided by the maximum possible 
produced number of kWh per year (power $\times$ 24 hours $\times$ 365 days), 
varies from more than 80\% for nuclear power, about 47\% from fossil fuels,
40\% for hydropower and about 20\% for other renewable energy sources.
These numbers indicate already the different ways of power plant operation.
In short, nuclear power plants provide the base load and gas and hydro power plants  
are great to provide power during peak load times. 
Wind and other renewable power plants, when available, contribute to the demand in such a way 
 that the use of the flexible gas and hydropower plants can be reduced.

Depending on the country, the time of the day and the period of the year, the electric power 
demand varies by about a factor of up to 2. The daily minimum load demand, a measure for the base load, 
is found during night at around 4:00 
and the maximum, the peak load, in the early evening between 17:00 to 19:30. 
The grid system has been optimized during the past years to adapt to those 
demand variations. Overcapacity in some regions and during different times of the day are used to compensate, 
as quickly as possible, to demand variations and potential power shortages in other regions and countries.

Nuclear power plants provide on average about 30\% of the yearly electric energy  
within the european system.
Due to the technical difficulties to regulate the electric energy production, 
nuclear power plants are ususally running always at 100\% capacity and 
provide therefore the base load energy.
However, especially in France a large nightly overproduction of electric energy still exists.
This overproduction is partially used for hydro power pump storage systems, 
which are found mainly in countries with high mountains. Another part of this excess electric energy  
is used for some more wasteful applications like 
too much light at night and perhaps the now widely spread standby mode of modern 
electric applications. In contrast to this base load, the peak load power is mostly produced by hydropower and 
gas fired electric power plants.

So far, the peak load demand during the summer time is much lower, due to the absence of heating, 
the smaller needs for electric lightening and (at least so far) little needs for climatization.
Most of the past record consumptions in Europe occured during the winter season and some 
particular ``cold" days, most likely to happen between December and January, 
excluding the end of the year holiday period.  

Especially due to electric heating, winter peak loads are very temperature dependent. 
One finds that countries which produce a large fraction of their overall electric energy 
with nuclear power plants, have a larger temperature dependent peak load
demand than those who have little or no nuclear power plants.   
The statistics, accumulated over many years, indicate that during the winter heating season 
about 1.7 GWe additional capacity per degree temperature drop is required in France\cite{lemonde}. 
This corresponds to about 27 Watt per person and degree. 
In comparison,  for Switzerland one finds that the corresponding per capita increase per degree of temperature drop 
is about a factor of three smaller, e.g. 
a total increase of  0.07 GWe/degree celsius corresponding to about 9 Watt/degree and per person. 

One reason is that the fast growth of nuclear 
power plants, 20-30 years ago, resulted in a large overcapacity and overproduction of 
electric power during night and in the summer and electric house heating was encouraged.
Thus, flexible and cheap electric heating systems, some with
overnight heat storage capacity, became very competitive and, at least in the short term,
cheaper, easier to use and cleaner than fossil fuel heating systems. 

Another more important aspect for the use of this overcapacity helped to improve 
the stability of the grid. The base load surplus electricity could be used 
to pump water uphill and release it again during higher load demand times. 
If one thinks only in terms of the total produced usable electric energy, this storage system is a substantial loss.
However, it appears that this solution is probably  
more efficient than the loss which happens if fossil fuel power plants are operated  
in a continuously changing on/off mode. 
Pump storage systems are obviously an economic benefit for countries with high mountains.
They can buy and store electric energy cheaply from the nightly overproduction  
and to sell the energy again during the day when the price is high.
As a result, a sophisticated system of such hydropower storage system 
was developed in regions and countries with high mountains like in Switzerland.

This comfortable period with sufficient amount of cheap electric energy 
transformed most people into some sort of drug 
addicts, who can not live even for a few hours without electric energy. In addition, 
the wish for better living standards corresponds thus also to more and more 
demand for electric energy.
Even though other options for better living can perhaps be imagined, 
the improved "living standards" of the past years corresponded to a yearly 
increase of  1-2\% of electric energy consumption and all hopes 
for further improvements are directly correlated with further annual growth rates 
of 1-2\%. 
However, this demand growth has literally eaten up the past spare power capacity 
and the expected continuing growth  
can not be satisfied for many more years without eventually increasing also 
the power plant capacity in Europe \cite{Axpo2020}. Thus, new power plants are urgently needed to replace 
aging ones and to enlarge the total capacity such that the next round of wished growth can continue. 

Depending on the assumptions about the future growth of electric energy demand 
and the required phase out of older power plants, either nuclear or conventional coal based, 
estimates for 2020 indicate that a new power plant capacity 
between 200 GWe (without demand growth) and 300 GWe (assuming a 1.5\% growth per year)  
is needed \cite{Axpo2020}. Similar numbers are estimated within the 
latest UCTE system adequacy outlook report for the years 2015 to 2020 \cite{ucte1}.
Within this report it is stated that todays net capacity is about 650 GWe. 
For the year 2015 a net capacity between 734-775 GWe, increasing further to 743-822 GWe by 2020,
is envisaged.

Even though these numbers include perhaps some generous rounding, 
they indicate the size of the problem for the european electric grid. Whenever this problem 
is discussed in the various european countries and organizations, the 
proposed solutions are more nuclear reactors, more gas and coal based CO$_{2}$ producing power plants
or more renewables like wind mills 
and mixtures of the three. As will be explained in the next section, it appears 
that none of the further ``growth"  proposals can be realized in sufficient quantities.

\subsubsection{The future nuclear option}

The so called {\bf nuclear option}, e.g. the building of new nuclear power plants,  is considered 
by some politicians, by some ``ivory tower" scientists and the ones who make their living from 
constructing these power plants the preferred solution. 
However, for various reasons, some countries like Germany have 
decided to exclude the construction of new nuclear power plants and even to 
terminate the use of nuclear power \cite{germanKKW}. The german nuclear capacity, 
currently  about 20 GWe, will be ``retired" during the next 15 years. 
Despite this praised or hated decision from the past government in 2001, this nuclear phase out program 
envisages to terminate each reactor, when its original planned retirement age of about 35 years is reached. 

In addition, the european nuclear option faces so far more problems than it offers solutions. 
First of all and within the european ``free" market, currently only one nuclear power plant option exists, the 1.6 GWe EPR reactor from Areva/Siemens
\cite{eprareva}. One such reactor is currently under construction in Finland.  
This reactor was supposed to be an example for a new generation of nuclear power plants in Europe. 
However this projects starts to become a serious embarrassment for 
the nuclear lobby. Reality shows, only two years after construction started,
that the promised 5 year construction time can not be achieved and that at least 7 years are required
\cite{eprfinland}. 
Another construction of an EPR reactor has ``started" a few month ago 
in Flamanville, France. It is currently expected that this reactor will be ready for commissioning 
and grid connection by 2012~\cite{Flamanville}.
We will soon know if the ambitious construction time of 54 month can be achieved.

So far, no other (public) decisions have been made, neither about additional reactors nor about the 
replacement problem for many european reactors, which will reach the retirement age during the next 10-15 years.
The future uranium requirements, in general planned over a long time,  
estimated by the EURATOM agency,   
indicate indirectly that the future nuclear power capacity in Europe will be reduced 
during the next 20 years by roughly 1/3 \cite{euratom}.
The document states that the currently required 21700 tons (for the EU25) of uranium per year
will be reduced to 17900 tons by 2016 and further to 14000 tons 
by 2026, a reduction of almost 8000 tons. 

To run a 1 GWe nuclear reactor for an entire year, one needs the equivalent of about 170 tons of natural 
uranium\footnote{For the first load of a new reactor about a factor of three larger amounts of uranium are required.}
Consequently one finds that todays 20 GWe German nuclear power plants require about 3300 tons uranium per year.
Thus, the estimated EURATOM reduction of 8000 tons by 2026 assumes that additional 20-30 GWe nuclear power capacity will be terminated 
during the next 15 years. 

This EURATOM estimate about reduced uranium requirements, translated into a reduced nuclear capacity, 
seems for now to be unknown or in contradiction with the latest UCTE system adequacy 
outlook. This study assumes that a ``retirement" of old nuclear reactors will only happen in Germany as 
documented by the expected reduction from todays capacity of about 112 GWe to 98.1 GWe by 2020.

Thus even the most radical nuclear enthusiasts, who 
ignore potential problems with (1) uranium supplies, with (2) the aging nuclear power plants, 
with (3) todays and tomorrows radical opposition against nuclear power,  
should acknowledge that their own pro nuclear EURATOM organization 
estimates a 30-40\% nuclear power reduction during the next 15-20 years. 

As decisions about site finding for new nuclear plants takes a long time 
and at least another 5 years are needed to really construct
new nuclear power plants, it seems likely that the contribution from 
nuclear fission energy to the european electric energy mix will decrease during the coming 10-15 years.

Another small problem for the nuclear power option is the acknowledged 
fact that nuclear power plants are always operated at 100\% power. 
Consequently, the additional nuclear capacity it not the optimal solution for the existing peak load problems as  
long as no additional storage capacity is constructed and or some larger electric energy use during base load times can be found.

One might simply conclude for now, that the most realistic perspective for the future of nuclear energy 
in Europe is somewhere between the 15\% reduction as estimated in the UCTE report and between 30-40\% (about 30-40 GWe) reduction 
as estimated indirectly from the EURATOM uranium requirement report.
As nuclear power plants can not function without uranium
one might think that the true power capacity numbers are perhaps better estimated with the EURATOM report.

\subsubsection{More fossil fuel powered electric energy?}

The alternative option to build {\bf new coal or gas powered} and thus CO$_{2}$ producing 
power plants seems to be the option most liked by the power providers. 
The reason seems to be that gas fired power plants are 
presently the cheapest and best solution to provide power during peak load problems.
In fact, the UCTE report assumes that the fossil power plant capacity 
will increase from todays 332 GWe to 363.1 GWe or even 417.1 GWe by 2020~\cite{ucte1}.
Strange enough, the problem with CO$_{2}$ emissions is not even mentioned. 
As no references, about where and when these new fossil fuel power plants should 
be completed, are given one might think that this ``solution" corresponds more to the 
wish list of the power plant operators than to a real firm planning for new power plants.  

In any case, such new CO$_{2}$ emissions are not 
compatible with promises from the european countries to reduce CO$_{2}$ emissions 
by at least 20\% and in the year 2020 \cite{Barosojan08}.

In addition, the unclear supply situation for gas and perhaps also for coal 
does not increase the confidence in this fossil fuel power plant solution either.
It should be obvious that without sufficient domestic gas resources, 
Europe will depend more and more 
on gas imports from Russia and other far away countries. Consequently, the  
idea to increase the electric power production with fossil fuel power plants 
does not receive much enthusiasm either.  

\subsubsection{Electricity production with so called renewable energy.}

The third option is the so called {\bf renewable energy option}.
{\bf Hydropower} is already largely used in Europe 
and not much potential exists to increase it any further. 
{\bf Wind power} saw a huge capacity increase during the last years. 
About 2 GWe per year of new wind power capacity was installed yearly in Germany alone and
a further large increase of a few GWe per year can be expected during the next 10-15 years.
Unfortunately, 2 GWe is a small number compared to the required 50 GWe. 
As wind is not always blowing, the average 
availability factor reduces this realistically to an effective 0.4 GWe power or less \cite{windcapf}.  
It is also known that wind power is not really  
optimal to provide stability to a stressed peak load times of the electric grid \cite{renewablecap}.
Other renewable electric power sources, like solar and geothermal power plants contribute so far, and despite large \% growth rates, 
a total of only 1-2 GWe installed power. Their capacity factors are currently estimated to be 0.1 or less \cite{solaretc}.  
 
\subsubsection{The option to reduce our addiction to electric energy.}
 
The only remaining option,  which is rarely discussed, is to first stop the annual {\bf growth} of 
our electric energy consumption, especially during peak load times.  In a second step,   
one would {\bf reduce} our needs for electric energy in a ``controlled" way. 
This reduction option is for most decision makers and those who influence
a {\bf ``no option"}. 
In fact this solution is currently totally unacceptable and even unthinkable.

\subsection{Assumptions about available reliable power and the future of produced electric energy.}  

If one uses the net power capacity or even the roughly 30\% smaller figure for the reliable available 
capacity and integrates this number over a day and over the year,
it seems that in principle enough kWh of electric energy can be produced.
Consequently, most people are let to the belief that 
any supply problem lies far away in the future and that as a worst case 
scenario,  the future energy mix is perhaps a bit unclear. For example during the last 10 years 
the installed capacity of wind power in Germany had an impressive growth rate and reached up to 2 GWe per year.
The total wind power has now a capacity which already exceeds  
the capacity of the German nuclear power plants. However, taking the capacity factor into account 
one finds that integrated over the year, about a factor of five more wind power capacity in comparison with 
nuclear fission power is required to 
produce the same amount of kWh.
Furthermore, the real availability of wind power shows large fluctuations, 
which makes it almost impossible to provide enough safety for particular peak load demand fluctuations.

Another problem is related to the belief that the extraordinary reliability of the different types of
power plants, observed during the past 10-15 years, 
will remain unchanged despite their old age also during the coming 10-15 years. 
As this belief is based on a few assumptions, it might be useful to remember them:

\begin{itemize} 
\item Hydropower plants and the capacity will not be affected in any dramatic ways. 
This belief is perhaps doubtful as the  
effects from climate change will change the times of snow melting and longer draught periods should be expected. 
\item The import of gas and coal will remain as stable as in the past and the only imaginable 
consequences of shortages might be some price changes. As the price of electricity is 
still for most ``richer" Europeans relative cheap,  
most people do expect that sufficient resources will exist to even satisfy a further demand growth of 1-2\% 
at least for the next 10-15 years.  
Despite the ever growing dependence on gas and coal imports from 
so called unstable regions and from far away,  no supply interruptions are imagined for the future. 
\item The promised reductions of CO$_{2}$ emissions by about 20\% from the European
Union will not result in any reductions of the number of produced kWh from fossil fuel power plants. 
  \item Uranium supply from outside Europe will be sufficient 
for whatever path the nuclear option will take on a world wide scale.   
\item The installed electric power capacity, despite its aging,  
will remain as reliable as in the past.
\end{itemize}
 
Each of these assumptions should be discussed in great detail as almost any of them faced some 
contradictions during the past few years. For example, the extremely dry and hot summer during 2003 
reduced the amount of produced hydropower and the river temperature reached levels 
that cooling limits for some nuclear power plants forced them into a reduced energy production mode \cite{summer2003}.

The current political will to reduce CO$_{2}$ production in Europe 
might in fact be not much more than pre-election populist propaganda. 
In any case, a more consistent discussion about the future of fossil fuel power plants 
can be expected during the next few years.  

Another worrying point is related to the security of the gas supply and the gas pipelines from Russia.
Most people expect that the growing dependence of Western Europe will only lead to an increased 
and unwanted political pressure from Russia. A similar situation might exist also with the supply 
of uranium, for which only about 50\% is assured to come from countries with 
so called stable market conditions. 
Thus  the ``limitless" supply assumption for gas, uranium and perhaps coal for the countries within the european grid 
system, which requires the stability and {\it ``good will"} from far away countries, does not appear to be realistic.
Finally, everyone should think about the assumption that 
aging nuclear and fossil power plants will always function as reliable as in the past.  

It can be concluded that each of the above assumptions has some limited basis 
and that it might not be wise to accept them as a good guidance for the future reliability of the electric grid system.
It seems more realistic that the grid stability and grid stress will become more 
and more critical during the coming years. 
Most people who think about the situation and who influence 
political decisions about the electric grid and the coming problems acknowledge at least some parts of the 
above arguments. Nevertheless, the UCTE and most policy makers continue to assure the public 
that the overall kWh consumption with an increase of 1-2\% can go on for another few years.    
However, as will be discussed in the next section, this approach 
is far too optimistic as the real supply problem is related to the peak load demand 
especially during cold winter days.

\section{Winter Peak load and severe winter conditions} 

The demand during peak load times was growing during the past years and in many countries 
by about 2\% per year, while the reliable supply did not really increase in most countries.
For example, the reliable capacity from France, as estimated in the UCTE winter outlook reports 2007/8 \cite{UCTEWinter07}. 
was about 90 GWe in January, decreasing to about 80 GWe by end of March. This estimate is  
about 1 GWe lower than in the previous winter outlook report from 2006/7 \cite{UCTEWinter06}.  
As a result, the peak load demand in France under severe 
conditions\footnote{Severe conditions are assumed when the temperature drops by 5 degree 
below the long term average temperature, a situation expected for some days during 
every ``normal" winter \cite{uctesevere}.}  was found to be about equal 
to the total available reliable capacity. Similar critical conditions 
appear now also for several other countries assuming severe winter temperatures.  

It might thus be interesting to take a closer and critical look at the findings of 
this latest UCTE winter 2007/08 outlook report \cite{UCTEWinter07}.
Besides some general more assuring statements, this report    
explains that the critical situation is not really years away. It even states that already the peak load  
expected during a few cold days during the early weeks of 2008 would have already resulted in a very stressed system.
In fact, it seems that a really serious risks for blackouts during peak load times 
were only avoided because of the extremely mild 2007/8 winter.

For example, a few days with severe cold January days, happening simultaneously in France, Germany and Switzerland,
like the ones observed during the ``normally cold" winter 2005/06
would have probably been sufficient to stress the system beyond capacity. 
A critical situation can thus be expected for the winter period of the next few years and especially during 
the peak load time. This time varies only slightly for the different countries. Taking some large per capita consumer countries like
France, Germany, Italy and Switzerland, which also exchange large amount of electric energy, one finds that   
the largest use of electric energy happens from Monday to Thursday and roughly in the early evening between 17:00 and 20:00. 

For various reasons, electric heating is widely used in France. 
The statistics, accumulated over many years, indicate that during the winter heating season, 
about 1.7 GWe additional capacity is required per degree temperature drop. 
As a consequence, the french system observed daily 
record consumptions on four consecutive days from Monday the 17th to 20th of December 2007 at around 19:00.  
The observed records from Monday to Thursday were 88.96 GWe, 88.60 GWe, 88.21 GWe and 86.42 GWe,
all these days were up to 5 degrees colder than the ``average" temperatures ~\cite{francerecord}.

These power demands should be compared with previous winter early evening records, 
of 86.28 GWe (27.1.06), 86.25 GWe (25.1.07)\footnote{The winter 2006/7 was besides a few days around the 25th of January 
extremely warm.}, 86.02 GWe (28.2.05) and 84.70 GWe (26.1.05). 
 
The new peak demand record on the 17th of December was roughly 3 GWe higher than 
ever before. During this evening, the French power plants exported about 3.5 GWe and imported 
about 5.1 GWe. Thus a net power of  
about 1.6 GWe was imported from Germany and Switzerland, \cite{lemonde}, 
which still had some reserve capacity. As it is not difficult to imagine that this spare 
capacity might not be available if a real cold wave hits simultaneously France, Germany and Switzerland,
one might wonder what is currently more likely: some ``voluntary" power reductions or 
large blackouts.

In any case, it is obvious that such a cold wave would not only bring 
severe stress to the French system but also to the European 
system as countries like Belgium, Great Britain and Switzerland are more or less 
dependent on the excess grid capacity from France.
The UCTE report claims that the weeks 2-4 of of 2008 represented a particular sensitive 
period for the French and thus the European grid system. Towards the end of the winter 
the requirements for heating and electric light especially during peak load time start dropping
and the situation will become less stressed. However   
the ``available reliable capacity will also be reduced from about 90 GWe in early January, to about 85 GWe 
by end of February and decreasing further 
to about 80 GWe by the end of March. 
For example, even during the exceptionally warm winter 2006/7, a few cold days were found to be 
below the long term average and the last year's winter peak load record 
in France was found on January 25 of 2007. 
In 2005, some cold days occurred even at the end of February, 
such that a peak load record of 86 GWe was observed on the 28th of February at 19:15.
It seems that the winter 2007/8 was again not cold enough, 
besides a few days in December, to really test the UCTE calculations. 
In any case, one should not relax completely before the next winter, as 
some unplanned capability loss or transmission line interruption due to technical failures, 
might always happen. In summary, the above numbers demonstrate that the critical peak load stress situation, 
is not years away but only a few days of severe winter weather in France and its neighboring countries.

\section{Measures to reduce the Grid Stress} 

The normal market reaction to shortages is obviously to increase the price of the particular 
good. However, it should be clear that demand regulation with electricity pricing and 
for smaller users can currently not be achieved on a short term notice. 
In absence of a strong centralized decision making system, small regional 
voluntary power outages, to reduce the peak load demand,
seem to be a difficult option, especially without sufficient preparation time.

Thus, possible nation wide preparations for short time power reduction with small inconveniences 
during the early evening peak load times and for a few particular cold days during coming winter periods 
might be a simple solution to reduce the risk  of country wide blackouts with all its nasty consequences.

What short term measures could be done? It is easy to imagine that action can be taken 
either nation wide or by individuals.
It is unlikely that enough idealistic behavior exists on a voluntary and unplanned basis, which would  
reduce consumption in such a way that the non idealistic people can simply continue their normal energy use. 
A potential alternative would be that some real political leadership shows up. 
Such leaders would explain the short term blackout risks during cold days
to a nation wide audience, followed by the demand 
to follow a few simple measures like the ones listed below:

\begin{itemize}
\item
Inform the public about any particular short term critical situation of the electric power grid and  
the potentially ugly consequences of blackouts. 
\item Remind the population 
that the critical peak load situation happens in the early evening between 17:00 and 20:00
and perhaps in the morning from 7:00 to 9:00 and around noon.  
\item Explain that already a slightly reduced usage of electric heaters (1-2 kWatt)
and various kitchen equipments during peak load time could make a sufficient difference 
to avoid blackouts. 
\end{itemize}  

The above measures could already be sufficient to reduce the risk for 
blackouts during the next few winter seasons.
However, real leadership would be required to explain how it was possible 
that no warnings have been given before and that no long term solution 
to the energy problem, besides the need to reduce energy consumption, exists.  
Consequently, a long term strategy about the energy problem
needs to be developed, following an open discussion with all members of the society.

The energy supply problem should be explained in great detail, 
following the points given below:
\begin{itemize}
\item Explain that the short term solution will not be sufficient in the longer term  
because: 
\begin{itemize}
\item It is impossible to replace the aging fossil and nuclear power plants fast enough;
\item Oil, gas, coal and uranium resources within Europe are already  
largely depleted; 
\item Resource rich countries, which currently export the gas, coal, oil and uranium for our power plants
are seeing a strong internal demand growth. It seems that countries like Russia 
will, within a few years, face a situation that they can either satisfy 
their internal needs or continue to export growing amounts of their energy resources.
The situation might soon reach the point where the leaders in Russia and other 
energy resource rich countries have to make unpopular decisions 
to reduce energy consumption within the country, while supporting the perhaps 
wasteful energy use in the richer countries of Western Europe. 
It seems rather unlikely that such an unpopular export strategy will be accepted for long 
within a reemerging powerful country.  
\end{itemize}

As a consequence of this situation, the public should understand that the total amount of 
available electric power and also of other forms of energy 
within the Western Europe can not be maintained for long and that the decline will start soon.
Logically the demand, including day and night variations, must be reduced efficiently and by the same amount!  
\item As the stability of the electric grid is currently fundamental for 
our functioning every day life and for our large scale society,
the only solution to a non chaotic way into the future is to learn how to reduce our dependence for 
electricity, and for all other forms of energy, as soon and as much as possible.
\item During this perhaps long and painful transition period, one should first find a way to reduce 
the use of the mostly unneeded electric energy applications.
\item Applications like electric heaters should even be forbidden or at least highly taxed to help financing 
better isolation and passive heating methods.
\item New projects, with large requirements for electric energy 
should only be accepted if simultaneously some similar power reduction 
measures in older existing projects can be achieved. 
\item Households and companies which manage to reduce energy consumption without damage,
should be shown as examples.   
\item How the different societies within the European Union will organize this decline in energy use 
depends on how fast and how radical this transition might be. 
But for sure, a rational approach towards a lower energy society, 
where the majority of its members can find a satisfactory way of living,  
requires something like a U-turn in essentially all areas of todays {\it ``way of live"}.
\end{itemize}

Unfortunately, it seems to be extremely unlikely that such political measures 
can be taken during the following months and years. The current political leadership 
should thus eventually be taken responsible for the consequences of any avoidable blackouts. 

Consequently, some personal blackout preparation might be as valuable 
as an assurance against other accidents. The simple minded examples 
below might help to reduce individual hardships during times of the severe grid stress
when the probability for blackouts will become larger and larger.  

\begin{itemize}
\item Understand the personal dependence on electric energy especially during 
peak load times.  
\item Avoid to take elevators, or choose the people which might be good company, 
especially during peak load times and during very cold days.
Likewise one should avoid all installations where it might be extremely unpleasant to be 
stuck. 
\item Care should be taken with automatic devices like window shutters. It might not be possible 
during longer blackouts to open them again. 
\item If cooking is done entirely on electricity, a safety camping gas kit might help 
for short term cooking problems.   
\item A collection of candles might be a good and even romantic idea.
\end{itemize}

However, none of these simple suggestions will prevent the ever increasing stress for the electric 
grid. Depending on the personal views about the energy problem one might either ask 
and campaign to build more power stations, conventional or so called renewable 
or figure out a way for a good living with reduced dependence on all kinds of 
energy sources. It might always be a good idea to remember and think about the fact that 
humans, like us, lived through good and bad times for about 100000 years without even knowing about electric energy, cars and planes. 
  
\section{Summary}

During the last 30 years a  fantastically stable european electric energy grid system 
has been established. Most citizens and companies of the UCTE member state countries enjoyed 
a situation, where any required amount of electric energy came with relatively small costs  
simply out of the plug. Demand fluctuations could be managed almost without problems 
and at any time of the day and the year. As a result, more and more 
electric applications were introduced to make everyday life ``simpler" and more ``enjoyable".
Even many obviously wasteful and sometimes decadent uses of electric energy became possible. 

The simplest extrapolation for the next few years follows the idea that the future 
will be like the past and electric energy demand and supply will thus continue to grow   
with 1-2\% per year. However, as it is obvious that new supply requires at some point 
new power plants of whatever type, reality seems to look different than the simple minded 
extrapolation into the future. The various data from the electric power production 
companies within the different European countries and even the optimistic 
extrapolations from the UCTE for the European 
grid, indicate that electricity supply gaps during the coming years, especially during the 
winter periods, are becoming more and more evident.

However, as with icebergs, the largest problems are normally associated with the parts hidden below the surface.
In contrast to the longer term supply gaps, the record peak load demands, observed in 
France and other countries on the evening of the 17th December 2007 indicate 
that the power capacity shortage problem is much larger than generally acknowledged.
In fact, it turns out that already and for the first time, France was faced with some 
shortage of electric energy if some colder than usual 
winter days would have happened during the early weeks of 2008. 
While the weather ``gods" helped again during the last winter and even with the observed 
global warming effects, the next winter seasons will come for sure.  
Thus, a further increase of the yearly peak power demand of 1-2 GWe, or about 2\% per year, in France alone can not 
for long be satisfied without building new power plants. 

As it takes at least five years or more to construct new conventional or nuclear power plants and 
as the future for secure imports of various energy resources looks more and more doubtful
and because of the CO$_{2}$ problem,  
not a single conventional technological solution for this supply gap seems to exist.
The largely unreliable power capacity associated with wind and hypothetical future 
solar power plants seems at best to limit the need for very drastic overall energy reduction measures.
Thus, the only solution for the short term electric energy supply problem which will avoid blackouts is  
a government based information campaign which aims to reduce first the electric energy demand 
during the peak load times. 

However, even if this new policy might be sufficient in the short term, it seems to us that 
the more and more obvious resource limits in Europe and in the rest of the world will force 
all of us on the path to a society with lower and lower per capita energy use. 
Those, who try to learn from historical examples might find that the ``way to live" 
in densely populated cities with little or no electric energy does not look too encouraging. 

Nevertheless, one might think that those who prefer to be active participants in our societies journey into  
the future can still choose to some extend about realizing their picture of a future low energy society. 
It is perhaps wishful thinking that your picture looks more like one with a small scale society of people working together 
rather than a top down hierarchical society with its very egoistic members at the various levels.
      
\vspace{1.cm}
\noindent
{\bf \large Acknowledgments} \\
{\normalsize  \it{
I would like to thank the many friends and colleagues, who could not escape discussions 
with me about the more and more disturbing energy problem.
Even if most of these discussions seem mostly to end nowhere,
I believe that they all contributed to the shape of this analysis about the electricity grid in 
Europe and perhaps its implications for other regions of our planet. 
In particular I would like to thank W. Hintz, F. Spano and W. Tamblyn for the critical 
reading of this paper and their encouragement and suggestions on how to improve the text.   
I am looking forward for further constructive criticism about this article and about ideas on how 
``we the people"  not only in Europe can find a non chaotic and non dramatic path 
into the unavoidable lower and lower energy future.}
}


\begin{thebibliography}{99}
\bibitem{uctesevere} {For France the severe winter conditions are defined on page A42 in \cite{UCTEWinter07}.}
\end{thebibliography}
\end{document}